\documentclass[amsmath,10pt,nofootinbib,prl,twocolumn,superscriptaddress, bibliography]{revtex4-1}
\usepackage{siunitx} 
\usepackage{microtype} 
\usepackage{amsmath}
\usepackage{amsbsy}
\usepackage{amssymb}
\usepackage{graphicx}
\usepackage{color}
\usepackage{physics}
\usepackage{soul}
\usepackage{bm}
\usepackage{array}
\usepackage{multirow}
\usepackage{lipsum}
\usepackage[capitalize]{cleveref}
\usepackage[normalem]{ulem}
\usepackage[normalem]{ulem} 
\usepackage[caption=false]{subfig}
\usepackage{xr}
\makeatletter
\newcommand*{\addFileDependency}[1]{
  \typeout{(#1)}
  \@addtofilelist{#1}
  \IfFileExists{#1}{}{\typeout{No file #1.}}
}
\makeatother

\newcommand*{\myexternaldocument}[1]{%
    \externaldocument{#1}%
    \addFileDependency{#1.tex}%
    \addFileDependency{#1.aux}%
}

\myexternaldocument{supplement}

\usepackage{verbatim}
\usepackage[bottom]{footmisc}

\makeatletter
\newcommand{\quickwordcount}[1]{%
  \immediate\write18{texcount -1 -sum -merge #1.tex > #1-words}%
  \immediate\openin\somefile=#1-words%
  \read\somefile to \@@localdummy%
  \immediate\closein\somefile%
  \setcounter{wordcounter}{\@@localdummy}%
  \@@localdummy%
}
\makeatother

\usepackage{natbib}

\begin{document}

\title{Thermodynamic Control of Activity Patterns in Cytoskeletal Networks}

\author{Alexandra Lamtyugina$^*$}
\affiliation{Department of Chemistry, University of Chicago, Chicago, IL 60637}

\author{Yuqing Qiu$^*$}
\affiliation{Department of Chemistry, University of Chicago, Chicago, IL 60637}
\affiliation{James Franck Institute, University of Chicago, Chicago, IL 60637}

\author{Étienne Fodor}
\affiliation{Department of Physics and Materials Science, University of Luxembourg, L-1511 Luxembourg}

\author{Aaron R.\ Dinner}
\affiliation{Department of Chemistry, University of Chicago, Chicago, IL 60637}
\affiliation{James Franck Institute, University of Chicago, Chicago, IL 60637}

\author{Suriyanarayanan Vaikuntanathan$^\dagger$}
\affiliation{Department of Chemistry, University of Chicago, Chicago, IL 60637}
\affiliation{James Franck Institute, University of Chicago, Chicago, IL 60637}


\begin{abstract}

Biological materials, such as the actin cytoskeleton, exhibit remarkable structural adaptability to various external stimuli by consuming different amounts of energy. In this work, we use methods from large deviation theory to identify a thermodynamic control principle for structural transitions in a model cytoskeletal network. Specifically, we demonstrate that biasing the dynamics with respect to the work done by nonequilibrium components effectively renormalizes the interaction strength between such components, which can eventually result in a morphological transition. Our work demonstrates how a thermodynamic quantity can be used to renormalize effective interactions, which in turn can tune structure in a predictable manner, suggesting a thermodynamic principle for the control of cytoskeletal structure and dynamics. 


\end{abstract}


\maketitle

\begingroup\renewcommand\thefootnote{$^*$}
\footnotetext{These authors contributed equally to this work.}
\endgroup
\begingroup\renewcommand\thefootnote{$^\dagger$}
\footnotetext{Corresponding author. Email: svaikunt$@$uchicago.edu}
\endgroup


The actin cytoskeleton is a paradigmatic example of an adaptive biomaterial that regulates important biophysical properties of the cell, such as its structural integrity, motility, and signaling, by adopting various non-equilibrium morphologies~\cite{Stam_Freedman_Banerjee_Weirich_Dinner_Gardel_2017,freedman2018nonequilibrium,schwarz2012united}. While there have been many efforts to unravel the driving forces responsible for sustaining many of these structures~\cite{koster2016actomyosin, weirich2019self, zhang2021spatiotemporal, Stam_Freedman_Banerjee_Weirich_Dinner_Gardel_2017,qiu2021strong,koster2016actomyosin,zhang2021spatiotemporal,ross_controlling_2019,Lemma_Mitchell_Subramanian_Needleman_Dogic_2021}, a clear thermodynamic understanding of the underlying principles governing their adapative properties has remained elusive~\cite{yang2021physical}. Here, using tools from large deviation theories~\cite{Tociu_Fodor_Nemoto_Vaikuntanathan_2019}, we provide evidence that a nonequilibrium thermodynamic control framework can indeed predict and rationalize adaptive structural transitions in cytoskeletal networks. Specifically, the central question motivating our work is: Can we predict how a cytoskeletal network adapts its structure to external conditions ({\it e.g.}, conditions requiring the formation of a contractile bundle) by controlling its energy budget?

\begin{figure*}[t]
\centering
\includegraphics[width=\linewidth]{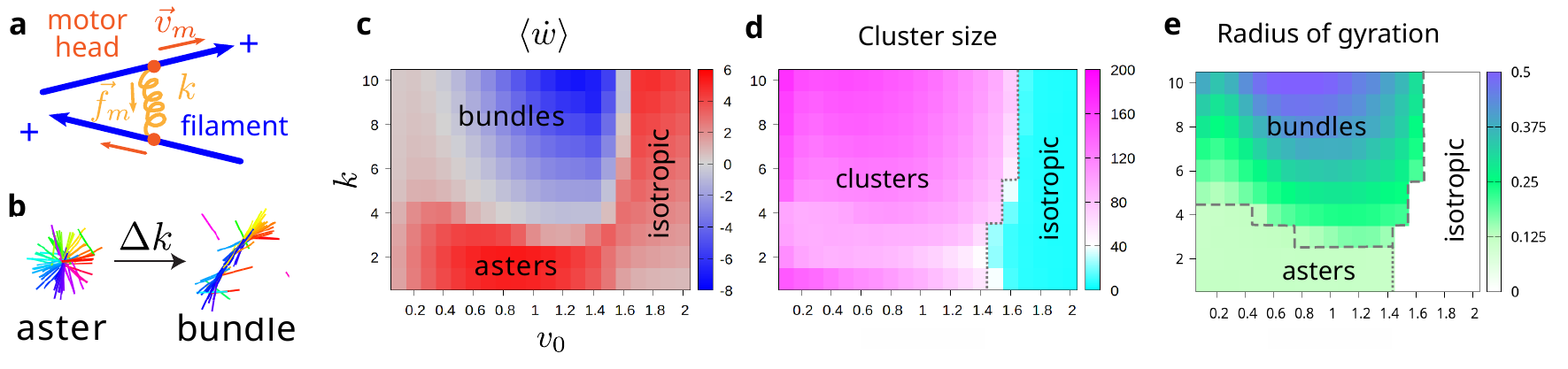}
\caption{
Structural transition between asters and bundles.
(a)~Schematic of two filaments (blue) connected by a motor (orange). The motor is modeled as a Hookean spring with rigidity $k$. The motor force $|\vec{f}_m|$ is proportional to motor rigidity $k$. Each motor head (dark orange) binds to one filament and moves towards the barbed ($+$) end with velocity $\vec{v}_m$ (\cref{eq:velocity-k}). When the motor is bound to two filaments, the rate of work done by the motor spring ($\Dot w $) is computed as the sum of $\vec{f}_m \cdot \vec{v}_m $ over the two motor heads.
(b)~Tuning $k$ induces the structural transition between asters and bundles.
(c)~Color map of $\langle \Dot{w} \rangle$. Asters and bundles are respectively associated with positive and negative values of $\langle \Dot{w} \rangle$. 
(d)~Color map of the largest cluster size. 
A system with no cluster larger than $40$ is considered to be isotropic. 
(e)~Color map of radius of gyration $R_g$ of the largest cluster. The boundary between bundle and aster is $R_{g} = 0.125$. The boundary of isotropic is the same as in (d). 
}
\label{fig1:unbiased}
\end{figure*}

To answer this question, we introduce a model that resembles \textit{in vitro} biomaterials consisting of actin filaments and molecular motors~(\cref{fig1:unbiased}(a)), and exhibits two well-known phases of such assemblies: asters and bundles~(\cref{fig1:unbiased}(b))~\cite{koster2016actomyosin,freedman2018nonequilibrium}. The core of the bundles is composed of anti-parallel actin strands resembling morphology found in stress fibers and cytokinetic rings \cite{Wollrab_2016, Reymann_2012}. The transition between these two states can be achieved in our model by modulating a material parameter related to the motor stiffness. Our main result shows how, by controlling the statistics of the rate of work done by the motors, the cytoskeletal network can transition between asters and bundles, thus generating configurations characteristic of different microscopic material properties. Importantly, this transition is now achieved even when the microscopic makeup of the cytoskeletal material (\textit{i.e.}, motor stiffness, motor speed, filament concentrations) are all held fixed. 

We obtain our results by building on recent theoretical studies~\cite{Cagneta2017, Nemoto2019, Tociu_Fodor_Nemoto_Vaikuntanathan_2019, Limmer2021entropy,das2021variational}, based on large deviation theory~\cite{Touchette_2009, jack2020} and stochastic thermodynamics~\cite{Seifert2012, fodor2021irreversibility,das2021variational} and applying them to our model actomyosin system. The framework of large deviation theory provides a convenient way to control the rate of work by applying a dynamical bias
to an ensemble of {\it trajectories}, namely a series of time realizations for the coordinates of motors and filaments. Combining detailed and specialized simulations with phenomenological theory, we reveal that the configurations generated with such a dynamical bias resemble those that would have been generated with a specific renormalization of the microscopic material properties. Specifically, for the regimes investigated here, given a cytoskeletal biomaterial composed of filaments and motors with a set stiffness and biochemical makeup, we show how  configurations characteristic of different values of motor stiffness can be accessed by simply modulating the statistics of the work done by the motors.

\begin{figure*}[tbp]
\begin{center}
    \includegraphics[width=\linewidth]{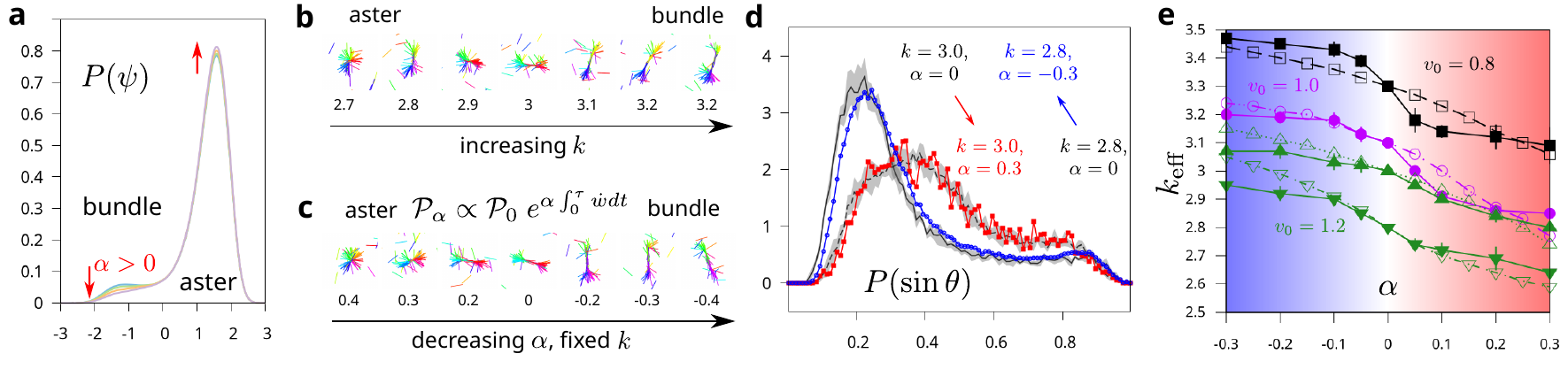}
    \caption{
Dynamical bias effectively renormalizes the motor rigidity $k$.
    (a)~Probability distribution obtained by biasing the dynamics of a given order parameter $\psi$ (with free energy $\mathcal F(\psi)= -a \psi + b \psi^2 - c \psi^3+ d \psi^4$) with respect to $\psi$. Parameters: ${a = 4d = -b = -30 c = 1}$. Biasing parameter: $\alpha = 0$~(unbiased dynamics, blue), $0.005$~(green), $0.1$~(orange) and $0.3$~(purple).
    (b)~Snapshots of unbiased simulations with changing motor rigidity $k$. 
    (c)~Snapshots of simulations from biased simulations
    with fixed motor rigidity $k=3$. 
    (d)~The statistics of structure from biased dynamics matches with that from an unbiased simulations at a different $k$. The order parameter $\sin \theta$ is calculated from angles between neighboring filaments and averaged across nearest neighbors. Matching the distribution of $\sin \theta$ from biased (black lines with the error bar shown with gray area) and unbiased simulations (red and blue) results in defining an effective rigidity $k_\text{eff}$.
    (e)~Effective rigidity $k_\text{eff}$ as a function of bias parameter $\alpha$ at $v_0 = 0.8$~(black), $1.0$~(magenta), and $1.2$~(green). The two green curves correspond to $k=3$ and $2.8$ analyzed for $v_0=1.2$. Filled points are obtained by matching structures from biased and unbiased simulations. Hollow points are analytical predictions derived from the two-state model (\cref{eq:lna}). 
    }
    \label{fig2:biased}
\end{center}
\end{figure*}

Our results suggest that controlling the rate of work, which could be achieved in practice by changing the consumption of chemical fuel~\cite{wu2021single,Speck2014}, can be regarded as a basic design principle for the development of an adaptive biomaterial. Below, we first introduce the microscopic coarse-grained description of actomyosin networks that we use in this paper. Next, we introduce tools of large deviation theory such as trajectory biasing, that allow us to probe the response of the system as the statistics of the work done by the motor are tuned. Finally, our main results in Fig.~\ref{fig2:biased} demonstrate how a biomaterial can access different classes of configurations, even though its microscopic makeup remains the same, when tuning the statistics of the work done by molecular motors.


Inspired by the rich phase diagram exhibited by actomyosin systems both \textit{in vivo}~\cite{Agarwal_Zaidel-Bar_2019} and \textit{in vitro}~\cite{Huber_Strehle_Kas_2012,koster2016actomyosin,freedman2018nonequilibrium}, we study the organization of short polar filaments connected by molecular motors using a coarse-grained platform, Cytosim~\cite{Cytosim}. Actin filaments and motors are, respectively, modelled as semi-flexible polymers and Hookean springs with filament binding sites at the two ends~(\cref{fig1:unbiased}(a)). Each motor head can bind to a filament and walk along its length towards the barbed end. When both motor heads are bound, the spring exerts a force $\vec{f}_m$ on the motor head in the direction pointing from the motor head to the center of the motor. The magnitude of $\vec{f}_m$ is determined by the motor rigidity, $k$, and the length of the spring, $l$: $|\vec{f}_m| = k l$. This force then modulates the loaded speed of each motor head $v_{m}$ as~\cite{Cytosim}:
\begin{equation}\label{eq:velocity-k}
\begin{split}
v_{m} = v_{0} (1+ \vec{f}_m \cdot \hat{d}/f_{0}),
\end{split}
\end{equation}
where $\hat{d}$ is the unit vector pointing from the motor head to the barbed end of the filament, $f_{0}$ and $v_0$ are friction force and velocity constants, respectively. As the motor heads walk along the filaments, they transmit the forces originating from the motor springs, and in response, the actin filaments can assemble into specific structures.

The phase diagram obtained by tuning the motor rigidity~$k$ and the motor unloaded velocity~$v_{0}$ is described in \cref{fig1:unbiased}(c-e). We characterize various regimes by calculating the size of the largest cluster of the filament-motor
cluster~(\cref{fig1:unbiased}(d)), and its radius of gyration $R_g$~(\cref{fig1:unbiased}(e), Sec.~S1C). Values of $R_g$ larger than half the length of a single filament indicate elongated bundle-like structures. Long bundles form at large $k$, and decreasing $k$ reduces $R_g$ until it reaches a plateau value corresponding to a radial arrangement of filaments, namely asters (Figs.~S1 and S2). This trend is consistent across different values of $v_0$ (Fig.~S2). Neither bundles nor asters form when the unloaded motor velocity $v_0$ exceeds a critical value, in which case a diffuse isotropic phase is observed. Our asters and bundles share core characteristics with those observed in experiments, such as the antiparallel alignment of filaments in sarcomeric bundles and stress fibers~\cite{Cramer_Siebert_Mitchison_1997} and clusters of radial filaments {\it in vitro}~\cite{Das_Bhat_Sknepnek_Koster_Mayor_Rao_2019,Stam_Freedman_Banerjee_Weirich_Dinner_Gardel_2017} (Fig.~S1). 

The rate of work due to the relative motion of the motor on the actin filament is defined as
\begin{equation}
    \label{eq:wDot}
    \Dot w = \sum_m \vec{f}_m \cdot \vec{v}_m,
\end{equation}
where $m$ runs over the number of motor heads~(\cref{fig1:unbiased}(a)). In what follows, we focus on the range of $v_0$ where the bundle-aster transition occurs, which is associated with a change of sign of the average rate of work $\langle \dot w\rangle$ (\cref{fig1:unbiased}(c)). In this regime, we aim to demonstrate that the transitions and structural changes that can be achieved by modulating the motor stiffness can equivalently be achieved, even when the motor stiffness and other material properties are held fixed, by modulating the statistics of $\dot{w}$ using tools from large deviation theory.

Specifically, we bias the trajectories generated in our simulations according to the rate of work (using cloning algorithm~\cite{tailleur2007probing, Nemoto_Bouchet_Jack_Lecomte_2016}, see Sec.~S1E) such that the probability of biased trajectories reads
\begin{equation}
    \label{eq:biased-probability}
    \mathcal{P}_\alpha \propto \mathcal{P}_0 \;e^{\alpha \int_0^\tau \dot{w} dt},
\end{equation}
where $\mathcal{P}_0$ is the probability of the trajectory in the absence of biasing, and $\tau$ the time duration of the trajectories. The parameter $\alpha$ tunes the strength and direction of the bias. In practice, trajectory biasing against the rate of work $\dot w$ can be considered as a way to probe the response of the system as the statistics of $\dot{w}$ are tuned~\cite{jack2020}. 
Besides, the specific choice of exponential reweighting in~\cref{eq:biased-probability} ensures that the distance between original and biased dynamics, as measured by the Kullback-Leibler divergence of their respective trajectory probabilities, is minimal~\cite{jack2020}. Below, we show that the configurations accessed when tuning the statistics of $\dot{w}$ resemble those that would have naturally emerged in a material with a renormalized motor rigidity $k$. 


Before proceeding to our numerical results, we first motivate how applying dynamical bias might impact system properties by considering a minimal phenomenological model of an actomyosin network~\cite{gowrishankar2016nonequilibrium}. Focusing on a simpler transition between an isotropic state and a state with asters, we show in Sec.~S2 that the dynamics of a relevant order parameter, $\psi$, may be described in terms of an effective free energetic landscape  $\mathcal F(\psi)= -a \psi + b \psi^2 - c \psi^3+ d \psi^4$, with $\{a,b,c,d\}$ as phenomenological parameters. Importantly, the phenomenological terms $b$ and $c$ are modulated by microscopic material constants such as $k$ and $v_0$, which can explain how the transition from the isotropic state to aster can be achieved by tuning the analogue of the motor rigidity in the phenomenological model~(Sec.~S2). Furthermore, we show how the application of a trajectory bias $e^{\alpha \int_0^\tau g(\psi)\, dt}$ results in dynamics that, at small noise, are equivalent to those generated by an effective free energy landscape~\cite{Nemoto_Bouchet_Jack_Lecomte_2016,tizon2019effective} but with renormalized phenomenological constants (Sec.~S2B). The probability distributions generated by biasing with various values of $\alpha$ are shown in \cref{fig2:biased}(a). 

This simple perturbative analysis reveals how an application of the bias can renormalize the phenomenological constants $b$ and $c$, effectively altering the motor rigidity. It also reveals how structural transitions obtained by tuning $k$ might also be achieved by tuning $\alpha$. This phenomenological model makes it reasonable to speculate that biasing the statistics of $\dot{w}$ in our coarse-grained simulations might effectively change the motor spring stiffness, opening up a different route to a structural transition. To confirm this intuition, we report in \cref{fig2:biased}(b-c) snapshots of the structure obtained in the biased dynamics without changing motor rigidity $k$ (Sec.~S1E), along with those of unbiased simulations with varying $k$. The similarity between these structural changes shows that biasing against $\dot w$ alone is indeed sufficient to induce the filament-motor system to move across the bundle-aster phase boundary. It also suggests that biasing as in~\cref{eq:biased-probability} might be effectively equivalent to modulating motor rigidity.

To quantify the effect of biasing on structure, we measure the relative alignment of filaments through the order parameter $\sin\theta$, where $\theta$ is the angle between neighboring filaments. We evaluate the order parameter by averaging $\sin\theta$ over the nearest neighbors for each filament (Sec.~S1F). The distribution, $P(\sin\theta)$, is shown for all filaments in the largest filament-motor cluster~(\cref{fig2:biased}(d)). For bundles at large $k$, the peak at $\sin\theta  \approx 0.25$ reflects parallel orientation of filaments. The distribution shifts towards larger values of $\sin \theta$ as $k$ decreases and the filaments rearrange into a radial aster. This order parameter $\sin\theta$ is related to the radius of gyration $R_g$ used to illustrate structural changes in ~\cref{fig1:unbiased}e, but they are not equivalent. $\sin\theta$ is more sensitive to the aster-bundle transition (Sec.~S1F). Therefore, we use $\sin\theta$ to quantify the effect of biaisng. For each biasing parameter $\alpha$, we define the effective motor rigidity $k_\text{eff}$ by matching the distribution $P(\sin\theta)$ measured in the biased dynamics with distributions obtained in the unbiased dynamics at $k=k_\text{eff}$. In practice, this matching is done by minimizing the divergence between these two distributions (Sec.~S1G), leading to a very good agreement between them (\cref{fig2:biased}(d)). Repeating this operation for different values of bias parameter $\alpha$ and unloaded motor velocities $v_0$, we obtain~\cref{fig2:biased}(e), which recapitulates the effect on the system structure of biasing the dynamics. This correspondence confirms that the effect of biasing against the rate of work is indeed fully accounted for as an effective change of motor rigidity, all other parameters being equal. 


Finally, since varying motor rigidity $k$ at fixed velocity $v_0$ leads to a transition between two distinct morphological states, asters and bundles, we aim at constructing a two-state model which, although minimal, is sufficient to rationalize quantitatively the effect of dynamical bias. We begin by assuming that the dynamics associated with the transition can be described by a master equation $\dot P = \boldsymbol{W}P$, where $P$ is the column vector with elements $\{P_\text{aster}, P_\text{bundle}\}$, and $\boldsymbol{W}$ is the transition rate matrix:
\begin{equation}
\label{eq:W}
    \boldsymbol{W} = 
    \begin{bmatrix}
        -R_{ab} & R_{ba} \\
        R_{ab} & -R_{ba} \\
    \end{bmatrix} .
\end{equation}
The entries $R_{ab}$ and $R_{ba}$ are meant to model the transition rates from aster to bundle and from bundle to aster, respectively. We express these rates using the Arrhenius law, $R_{ab} = A \exp[-\beta \varepsilon_{\text{bundle}}]$ and $R_{ba} = A \exp[-\beta \varepsilon_{\text{aster}}]$, where the energies of aster and bundle states are given by $\varepsilon_{\text{aster}}$ and $\varepsilon_{\text{bundle}}$, respectively, and $A$ is an Arrhenius prefactor. 
For convenience, we work in units such that $A=1$ and $\beta=1/(k_BT)=1$, and we set $\varepsilon_{\text{bundle}} = 0$.
To quantitatively connect this two-level picture with the simulation results of Cytosim, we relate energy levels to distributions by
\begin{equation}
  \varepsilon_{\text{aster}} = - \ln{\frac{P_{\text{aster}}}{1 - P_{\text{aster}}}},
\label{eq:epsilon_aster}
\end{equation}
where $P_{\text{aster}}$ is extracted from numerical data as $\int_{\sin\theta_c}^{1} P(\sin\theta)\,d\sin \theta$ with the choice $\sin\theta_c=0.6$, see \cref{fig2:biased}(d). 

The effect of applying a dynamical bias with respect to $\dot w$ is then recapitulated in terms of the master equation $\dot P^{(\alpha)} = \boldsymbol{W}^{(\alpha)} P^{(\alpha)}$. The transition matrix $\boldsymbol{W}^{(\alpha)}$ reads
\begin{equation}
    \label{eq:W_alpha}
    \boldsymbol{W}^{(\alpha)} =
    \begin{bmatrix}
        -R_{ab}+ \alpha \;\Dot w_{\text{aster}} & R_{ba} \\
        R_{ab} & -R_{ba}+ \alpha \;\Dot w_{\text{bundle}} \\
    \end{bmatrix} ,
\end{equation}
where $\Dot w_{\text{aster}}$ and $\Dot w_{\text{bundle}}$ are the rate of work for the aster and bundle states, respectively. The biased transition matrix is known as a ``tilted" matrix, and is constructed based on the principles of large deviation theory~\cite{Touchette_2009,Touchette_2012} (Sec.~S3A). We show that, to leading order in the bias $\alpha$ (Sec.~S3B), the effective energy level in biased dynamics $\varepsilon_{\text{aster}}^{(\alpha)}$ can be expressed as~\cite{das2021variational}
\begin{equation}
  \varepsilon_{\text{aster}}^{(\alpha)} \approx \varepsilon_{\text{aster}} - \alpha \frac{ (\Dot w_{\text{aster}} - \Dot w_{\text{bundle}})}{1+R_{ba}}.
\label{eq:lna}
\end{equation}
Eq.~\ref{eq:lna} hence predicts how the energy barriers may be modified due to biasing $\alpha$. This equation can be used to obtain a prediction for $k_{\rm eff}$ as a function of $\alpha$ by plugging the estimate of the modified barrier in Eq.~\ref{eq:epsilon_aster} and looking up the value of $k$ at which the estimate of $P_{\text{aster}}$ best matches the modified barrier height.  The quantity $\Dot w_{\text{aster}} - \Dot w_{\text{bundle}}$ in \cref{eq:lna} is best estimated from numerical values of $\langle\dot w\rangle$ close to the aster-bundle transition. To generalize this relation to regions away from the transition, we look to the meaning of $\dot{w}_{\rm aster}$ and $\dot{w}_{\rm bundle}$ in our two-state model. Specifically, these quantities are meant to denote the typical values of $\dot w$ in the regimes of high and low $\sin\theta$, respectively. Since away from this transition, the distribution $P(\sin\theta)$ is dominated by either the bundle or the aster phase, the difference in the typical values of $\dot{w}$ at high and low $\sin\theta$ is reduced. To effectively capture this reduction, we assume that,  to leading order, $\Dot w_{\text{aster}} - \Dot w_{\text{bundle}}$ is proportional to the slope of the $\Dot w$ versus $k$ curve, with the proportionality constant as a fitting parameter. This assumption, along with numerical estimates of $\varepsilon_{\text{aster}}$ for a range of $k$ values, enable us to predict how $k_{\text{eff}}$ changes with the biasing parameter $\alpha$ (Sec.~S3). Our prediction is in good agreement with $k_{\text{eff}}$ obtained numerically by directly matching the structure distributions taken from the biased and unbiased dynamics (\cref{fig2:biased}(e)). This agreement shows that our two-state model, although providing an over-simplified picture of the underlying dynamics, indeed captures the effective modulation of motor rigidity due to biasing the dynamics with respect to the rate of work.


%


A feature of probing the response of the system to $\dot{w}$ modulation in this manner (using the tools of large deviation theory) is that we do not provide any explicit protocol for how to perturb the energy consumption. We envision that experiments can be done by deploying active components such as light-sensitive motors~\cite{zhang2021spatiotemporal, ross_controlling_2019}, or by fueling the system with different ATP supplies~\cite{wu2021single,Speck2014}, which might provide a physical route for achieving such a perturbation. Our central results hence suggest a new route for the modulation of cytoskeletal material properties through the regulation of underlying energy consumption.

The ideas presented here are complimentary to existing hydroynamic treatments of actomyosin networks~\cite{Kruse_2005,Marchetti_2013}. These seminal works have shown how various actomyosin phases may be accessed by tuning phenomenological parameters, which in turn affects energy consumption (although in a way which can prove difficult to predict). Instead, our results reveal that {\it directly} tuning energy consumption, now in a much more predictable manner, is also a route to inducing structural transitions.
While we focus here on the connection between $\Dot{w}$ and network structure, our work may also provide a roadmap for understanding how cytoskeletal networks adapt to changing external stress conditions. Indeed, when the motor head velocity is a constant, $\Dot{w}$ is simply proportional to the force exerted by motors along the axis of the actin filament (\cref{eq:wDot}). In these regimes, tuning the statistics of $\Dot{w}$ is equivalent to tuning the axial forces exerted on the filaments. 
From a biological perspective, our work paves the way towards a thermodynamic understanding of the control principles regulating the cytoskeleton, to rationalize both how it adapts its structure to external cues~\cite{zhang2021spatiotemporal,ross_controlling_2019}, and how spontaneous flows can form as a result of internal activity~\cite{lecuit2007cell}. 

\acknowledgements{

 This work was mainly supported by a DOE BES Grant DE-SC0019765 through funding to SV, YQ. YQ was also supported by a Yen Fellowship. AL is supported by a NSF Graduate Research Fellowship DGE-1746045. EF was funded by the Luxembourg National Research Fund (FNR), grant reference 14389168. ARD acknowledges support from National Institutes of Health award R35 GM136381 and National Science Foundation award MCB 2201235.
 }
 
Additional references included in the SI are Refs.~\cite{Kurtzer_Sochat_Bauer_2017_Singularity,seabold2010statsmodels,Doob_1990}.
\bibliographystyle{unsrt}
\bibliography{references}

\newpage
\end{document}